%% file: main.tex
\documentclass{article}
\usepackage{spconf,amsmath,graphicx}

\usepackage{xcolor}         
\usepackage{amsmath}
\usepackage{booktabs} 
\usepackage{multirow, makecell}
\usepackage{threeparttable}
\usepackage{graphicx}
\usepackage{array}
\usepackage{caption}
\usepackage{amsmath,amssymb}

\newcommand{\sm}{supervised metric}
\newcommand{\smi}{\ensuremath{\mathrm{I}(Z;Y)}}
\newcommand{\um}{unsupervised metric}
\newcommand{\umi}{\ensuremath{\mathrm{I}(Z_a;Z_b)}}
\newcommand{\tnet}{\ensuremath{p_\psi}}
\newcommand{\rnet}{\ensuremath{f_\theta}}
\newcommand{\probe}{\ensuremath{q_\phi}}

\newcolumntype{H}{>{\setbox0=\hbox\bgroup}c<{\egroup}@{}}

\def\E{{\mathbb E}}

\title{REVISITING SELF-SUPERVISED LEARNING OF SPEECH REPRESENTATION \\
FROM A MUTUAL INFORMATION PERSPECTIVE}
\name{Alexander H. Liu$^*$\thanks{* Equal contribution.}$^1$ ~~~ Sung-Lin Yeh$^*$$^2$ ~~~ James R. Glass$^1$ }
\address{$^1$MIT CSAIL, $^2$University of Edinburgh, Informatics}
\begin{document}
%
\maketitle
\begin{abstract}
Existing studies on self-supervised speech representation learning have focused on developing new training methods and applying pre-trained models for different applications. However, the quality of these models is often measured by the performance of different downstream tasks. How well the representations access the information of interest is less studied. In this work, we take a closer look into existing self-supervised methods of speech from an information-theoretic perspective. We aim to develop metrics using mutual information to help practical problems such as model design and selection. We use linear probes to estimate the mutual information between the target information and learned representations, showing another insight into the accessibility to the target information from speech representations. Further, we explore the potential of evaluating representations in a self-supervised fashion, where we estimate the mutual information between different parts of the data without using any labels. Finally, we show that both supervised and unsupervised measures echo the performance of the models on layer-wise linear probing and speech recognition.

\end{abstract}

\input{tables/overview}
\input{sections/intro}

\input{sections/metric}

\input{sections/experiment}

\input{sections/conclusion}

\newpage
\small
\bibliographystyle{IEEEbib}
\bibliography{strings,refs}

\end{document}

%% file: tables/overview.tex
\begin{table*}[ht!]
\small
    \caption{Summary of prior self-supervised speech representation methods in two different categories: Autoregressive Predictive Coding (APC) and Masked Langauge Modeling (MLM). VQ stands for vector quantization; EMA stands for exponential moving average.}
    \label{tab:overview}
    \centering
    \begin{threeparttable}
    \begin{tabular}{lccccc}
    \toprule
    & Views  & \multirow{2}{*}{Choice of $Z_\text{target}$} & \multirow{2}{*}{Choice of $\tnet$} & \multirow{2}{*}{$\mathrm{H}(Z_\text{target})$ } & connection to \\
    & $X_a $/$ X_b $ &  & &  & $\mathrm{H}(Z_\text{target}|\rnet(X_a))$ \\
    \midrule
    \multicolumn{5}{l}{APC family} \\
    ~~APC~\cite{chung2020generative} / VQ-APC~\cite{chung2020vector} & \multirowcell{2}{past / future~} & spectrogram & identity matrix & intractable, fixed  & regression\\
    ~~Co-training~\cite{yeh2022autoregressive} &  & discrete & VQ & estimable$^\ddagger$ & cross-entropy\\
    \midrule
    \multicolumn{5}{l}{MLM family} \\
    ~~wav2vec 2.0~\cite{baevski2020wav2vec} & \multirowcell{3}{masked$^\dagger$~/\\unmasked} & discrete & VQ & estimable$^\ddagger$ & contrastive loss \\
    ~~HuBERT~\cite{hsu2021hubert}~/~WavLM~\cite{chen2022wavlm} &&  discrete & k-means assignment & estimable$^\ddagger$, fixed & cross-entropy  \\
    ~~DinoSR~\cite{liu2023dinosr} & &  discrete & EMA(\rnet) + VQ & estimable$^\ddagger$ & cross-entropy \\
    \bottomrule
    \end{tabular}
    \begin{tablenotes}
    \item{$\dagger$} \small{WavLM additionally introduced noise to generate $X_a$. $^\ddagger$ We consider $\mathrm{H}(Z_\text{target})$ estimable by sampling $Z_{\text{target}}$ from $\tnet(Z_{\text{target}}|X_b)$ to compute the empirical entropy.} 
    \end{tablenotes}
    \end{threeparttable}
    \vspace{-5pt}
\end{table*}
  

%% file: sections/intro.tex
\section{Introduction}

Estimating the amount of information encoded in learned representations has been an important research topic in speech representation learning.
A good estimation can not only offer a better view of designing training objectives, especially under a self-supervised paradigm but also help select models for the applications of interest. 
To measure to what extent the representations reveal specific information, several studies have adopted a phonetic-related linear probing protocol \cite{chung2020generative,yang2022autoregressive}.
Also, a collection of downstream tasks has been proposed to evaluate learned representations, including phone classification and speech recognition \cite{yang2021superb}. The accuracy obtained from a task is then believed to reflect the accessibility of representations to certain information.

Although the aforementioned approaches for measuring representations have been widely used, there are certain limitations. For example, a probing task is not formally measuring the ``information'' inherent in the representations but their accuracy on a task.
Another limitation is that the probing tasks all rely on labeled data.
Further, contextual speech representations learned from self-supervised models are actually trained to predict the context such as the future or masked frames \cite{baevski2020wav2vec,hsu2021hubert,chung2020generative,chung2020vector,yang2022autoregressive}. The current probing approaches, however, ask the classifier to do same-frame prediction.   
The mismatch between the training objectives and the evaluations makes it unclear whether the current measurements have properly reflected the information representations encoded. 

In this paper, we present an information-theoretic approach to assess the information contained in representations \cite{pimentel2020information,voita2020information}. We use mutual information (MI) to measure the relationship between representations and their targets, such as phonetic labels.
To evaluate how well representations capture the context, we propose that effective representations should exhibit higher MI between different parts of the input due to self-supervised training. To test this, we divide the input into two parts and estimate the MI between representations derived from different parts. This offers an unsupervised alternative to measure the learned representations.

Through extensive experiments, our findings reveal a strong correlation between unsupervised measures and supervised ones in phonetic-related probing. This correlation suggests the potential for probing representations without labeled data. Furthermore, we observe that models exhibiting higher MI in an unsupervised measure also exhibit superior performance in downstream speech recognition.

%% file: sections/metric.tex
\section{Measuring Self-supervised Models with Mutual Information}

\subsection{A Mutual Information Perspective of Self-supervised Methods on Speech}

Prior works have drawn the connections between self-supervised training objectives and maximizing mutual information (MI) \cite{oord2018representation,kong2019mutual,yang2022autoregressive,laina2022measuring}.
Following a similar vein, we consider self-supervised approaches as maximizing the MI between different parts of the input by dividing the input into two views, defined as $X_a$ and $X_b$. Specifically, we focus on the MI between $X_a$ and target variables $Z_\text{target}$ derived from $X_b$.
Formally,
\begin{align}
    \mathrm{I}(X_a;Z_\text{target}) &= \mathrm{H}(Z_\text{target}) - \mathrm{H}(Z_\text{target}|X_a), \label{eq:ssl_mi}
\end{align}
where a network $f_\theta$ is employed to model $p(Z_\text{target}|X_a)$.
Besides the different views of data, the choice of target variable $Z_\text{target}$ is perhaps the most significant difference between methods.
Some methods~\cite{chung2020generative,hsu2021hubert,chen2022wavlm} have propose to use pre-defined transformation $\tnet(X_b)=Z_\text{target}$ to derive target variable; other methods~\cite{baevski2020wav2vec,liu2023dinosr} have proposed to learn it jointly during training by introducing $\tnet(Z_\text{target}|X_b)$.
Table~\ref{tab:overview} provides some concrete examples.
Note that we only list open-sourced models in the table, there are more prior works~\cite{ling2020decoar,chung2021w2v,chiu2022self} in the field that are not covered.

Theoretically, measuring MI through Eq.~\ref{eq:ssl_mi} allows us to compare different SSL methods in speech, but in practice, it is infeasible due to some limitations.
For example, methods like APC~\cite{chung2020generative} and VQ-APC~\cite{chung2020vector} in Table~\ref{tab:overview}, Eq.~\ref{eq:ssl_mi} is intractable due to the unknown distributions in the equation \cite{mcallester2018information}.
Even in the case where Eq.~\ref{eq:ssl_mi} can be approximated (e.g., HuBERT~\cite{hsu2021hubert}), it is hard to fairly compare it to other methods due to the different definitions of $\tnet$ and $Z_\text{target}$.
This motivates us to bound MI in a tractable way that is invariant to the choice of $Z_\text{target}$ and $\tnet$ such that we can compare these self-supervised methods.

\subsection{Bounding MI with labeled data}

After the SSL stage, the representation $Z$ of data $X$ can be extracted with the pre-trained model $\rnet$, i.e., $Z = \rnet(X)$.
An intuitive way to compare the quality of speech representations $Z$ from different models is by examining the mutual information between $Z$ and specific \textit{target} $Y$ using labeled data (e.g., the underlying phone at the corresponding time), namely, $\mathrm{I}(Z;Y) = \mathrm{H}(Y) - \mathrm{H}(Y|Z)$.
While the entropy of target $\mathrm{H}(Y)$ is a constant (depending solely on the choice of target) that can be estimated with an empirical distribution, the metric itself is still intractable since we do not know the relation between $Z$ and $Y$.
Nevertheless, the target mutual information $\mathrm{I}(Z;Y)$ can be lower-bounded through upper-bounding the conditional entropy $\mathrm{H}(Y|Z)$ with an auxiliary prediction model $\probe(y|z)$.
More precisely, 

\begin{align}
    \smi &= \mathrm{H}(Y) - \E_{(y,z)\sim p(Y,Z)}\Big[-\log p(y|z)\Big] \label{eq:sup_mi} \\
    &= \mathrm{H}(Y) - \E_{p}\Big[-\log \probe(y|z) -\log \frac{p(y|z)}{\probe(y|z)} \Big] \\
    & \geq \mathrm{H}(Y) - \E_{p}\Big[-\log \probe(y|z)\Big] \label{eq:ce},
\end{align}
where $\E_{p}\big[\log\frac{p(y|z)}{\probe(y|z)}\big] = \E_{Z} D_\text{KL}(p(y|z) || \probe) > 0 $ leads to Eq.~\ref{eq:ce}.
In other words, we can estimate the lower bound of the desired mutual information by training the auxiliary prediction model $\probe(y|z)$ to approximate $p(y|z)$.
Note that this corresponds to \textit{probing tasks} in the literature~\cite{chung2020generative,yang2021superb} since the last term in Eq.~\ref{eq:ce} is cross-entropy loss. For example, in linear probing tasks, $\probe(y|z)$ is modeled by a linear layer.

\input{tables/main_result}

By using labeled data and Eq.~\ref{eq:ce}, we are able to establish a lower bound of mutual information $\mathrm{I}(Z;Y)$ that can be used as an intuitive metric to measure the quality of representation.
However, there are several downsides making this metric less ideal for selecting SSL models such as the need for labeled data and the narrow viewpoint from the choice of target.
These properties somewhat contradict the spirit of the SSL paradigm for learning a general model that can be applied to different tasks with minimum supervision.

\subsection{Bounding MI with unlabeled data}

In light of how self-supervised methods are designed, we propose to measure the mutual information between different views of the input instead of using labeled data.
For simplicity, here we use $Z_a=\rnet(X_a)$ to denote the representation extracted from one view and $Z_b=\rnet(X_b)$ from the other. 
Formally, we consider
\begin{align}
    \umi &= \mathrm{H}(Z_a) - \mathrm{H}(Z_a|Z_b).
\end{align}
Similar to Eq.~\ref{eq:sup_mi}, the mutual information is intractable since the underlying distributions of the variables are unknown.
To overcome the issue, we introduce a clustering function $f_\text{cluster}$ to quantize the representation $Z_b$. We can then approximate $\mathrm{H}(Z_b)$ with the empirical distribution and estimate a lower bound of the mutual information with
\begin{align}
    ~\umi \geq &~\mathrm{I}(Z_a;f_\text{cluster}(Z_b)) \label{eq:dp} \\
    = &~ \mathrm{H}(f_\text{cluster}(Z_b)) - \mathrm{H}(f_\text{cluster}(Z_b)|Z_a) \label{eq:km_mi} \\
    \geq &~ \mathrm{H}(f_\text{cluster}(Z_b))  - \E_{p}\Big[-\log \probe(f_\text{cluster}(Z_b)|Z_a)\Big], \label{eq:unsup_mi}
\end{align}
where Eq.~\ref{eq:dp} follows from the data-processing inequality and Eq.~\ref{eq:unsup_mi} can be derived in a similar way to Eq.~\ref{eq:ce}.
The key advantage of this approach is the lower bound can be estimated regardless of the choice of $Z_\text{target}$, making cross-method comparisons and checkpoint selections possible without labeled data as we show later in our experiments.

%% file: tables/main_result.tex
\begin{table*}[ht!]
    \small
    \begin{minipage}{0.60\textwidth}

    \caption{Results of MLM methods on LibriSpeech test-other.
    For downstream tasks, frozen feature results are taken from Speech processing Universal PERformance Benchmark~\cite{yang2021superb} and the 10-hour fine-tuning results are the decoding result with language model reported by each prior work. 
    } 
    \label{tab:mlm}
    \centering
    \begin{tabular}{lccccccc}
    \toprule
    & \multicolumn{4}{c}{mutual information lower-bound}  & \multicolumn{3}{c}{downstream tasks} \\
     \cmidrule(l{2pt}r{2pt}){2-5}\cmidrule{6-8}
    & \multicolumn{2}{c}{\smi~{\small {(Eq.~\ref{eq:ce})}} }  & \multicolumn{2}{c}{\umi~{\small {(Eq.~\ref{eq:unsup_mi})}} } & \multicolumn{2}{c}{frozen feature} & fine-tune  \\
    \cmidrule(l{2pt}r{2pt}){2-3}\cmidrule(l{2pt}r{2pt}){4-5}\cmidrule(l{2pt}r{2pt}){6-7}\cmidrule(l{2pt}r{2pt}){8-8}
    & logistic & MLP & logistic & MLP &  PER & WER & WER \\
    \midrule
    \multicolumn{3}{l}{\textsc{Base} models (12-layer)} \\
    ~wav2vec 2 & 3.50 & 3.52 & 1.13 & 1.08 &  5.74 & 6.43 & 9.5 \\
    ~HuBERT & 3.73 & 3.75 & 2.04 & 2.04 &  5.41 & 6.42 & 9.4 \\
    ~WavLM & 3.80 & 3.82 & 2.05 & 2.05 &  4.84 & 6.21 & 9.2 \\
    ~DinoSR & 3.83 & 3.89 & 2.10 & 2.15 &  3.21 & 4.71 & 7.6 \\
    \midrule
    \multicolumn{3}{l}{\textsc{Large} models (24-layer)} \\
    ~wav2vec 2 & 3.81 & 3.87 & 1.67 & 1.63 &  4.75 & 3.75 & 4.9 \\
    ~HuBERT & 3.85 & 3.90 &  2.08 & 2.05 & 3.53 & 3.62 & 4.6 \\
    ~WavLM & 3.87 & 3.94 & 2.40 & 2.36 &  3.06 & 3.44 & 4.6 \\
    \bottomrule
    \end{tabular}
    \end{minipage} 
      \begin{minipage}{0.40\textwidth}
    \captionsetup{width=\linewidth}
    \centerline{\includegraphics[width=\linewidth]{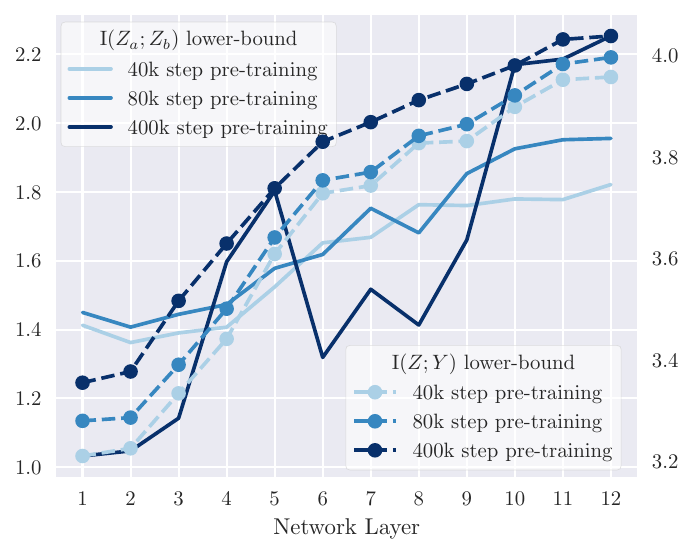}}
    \vspace{-3pt}
    \caption{Lower-bound dynamic w.r.t. different layers and pre-training steps of DinoSR. Y-axis on the left corresponded to \umi~lower-bound (bit); right corresponded to \smi~lower-bound (bit).}
    \label{fig:layer}
  \end{minipage} 
\end{table*}

%% file: sections/experiment.tex
\section{Experiments}

\noindent\textbf{Setup.} 
We use k-means as the clustering function $f_\text{cluster}$ to quantize the representation space with default 50 clusters.
For $\probe$, we test both logistic regression and multi-layer perceptron (MLP) with 3 layers, ReLU activation, and dropout.
Since our goal is to evaluate pre-trained self-supervised models, both $f_\text{cluster}$ and $\probe$ are trained on the clean dev set of LibriSpeech~\cite{panayotov2015librispeech} and used to estimate MI on test-clean/test-other for APC/MLM models respectively.
By default, we use k-means with 50 clusters for a max of 100 iterations, then optimize Eq~\ref{eq:unsup_mi} with a learning rate of 0.1 for 10 epochs.
For \smi~, we use force-aligned~\cite{mcauliffe2017montreal} phone sequences as the target $Y$.
For MLM models, we follow the same setup but adopt the noisy subsets.

For APC models, views $Z_a$ and $Z_b$ in Eq.~\ref{eq:ssl_mi} are generated by applying a time shift identical to the pre-training stage (60ms; 3 frames) to the representations.
For MLM models, the masked view is generated by masking the last 30 frames of every 40-frame (i.e., 75\% masking ratio), a more detailed discussion is provided later with results on different masking ratios.
Representations are extracted from the last layer unless otherwise specified.
All numbers reported are averaged over 5 runs with different random seeds, we find the variance across different runs negligible ($<$4e-4) for all cases.

\input{tables/apc}

\vspace{3pt}
\noindent\textbf{Bounding the MLM family.} 
We begin with results in Table~\ref{tab:mlm} on the MLM family that take masked/unmasked speech as the different views.
We compare both \sm~and \um~ against the downstream performance of each model.
For downstream performance, we consider the speech recognition performance from SUPERB~\cite{yang2021superb}.
Unsurprisingly, there is a strong connection between the \sm~ and downstream performance. Models with higher \smi~ provide representations with higher accessibility of phonetic information and benefit phonetic-related tasks.
On the other hand, a similar pattern can be observed with the \um~(despite not using any labeled data) where the increasing lower bound of \umi~ reflects stronger downstream performance.
This key result suggests that it is possible to evaluate self-supervised speech representations in a self-supervised manner.

\vspace{3pt}
\noindent\textbf{Impact of bounding conditional entropy with $\probe$.} Another observation from Table~\ref{tab:mlm} worth mentioning is the choice of $\probe$ actually have small impact to our estimated lower bounds.
The results are consistent between the two choices with slightly better estimation obtained via MLP in most cases.
While only two different options are tested in this paper, we note that exploring better options of $\probe$ is worthwhile in practice since they provide tighter lower bound~\cite{pimentel2020information}.

\vspace{3pt}
\noindent\textbf{Bounding the APC family.} 
In addition to MLM, we experiment with models trained on future prediction pre-training, in which the past and future are selected as different views.
As shown in Table~\ref{tab:apc}, 
the trends of supervised and unsupervised metrics do not completely align with the results of phoneme classification. Nonetheless,
the layer with the highest MI consistently corresponds to the best PER of the model.
This indicates the proposed metrics can be potentially applied to
layer selection for phonetic-related tasks.
Note that even though the number of clusters is fixed, \um~is not comparable to that of the MLM methods due to different model configurations.

\vspace{3pt}
\noindent\textbf{Layer-wise analysis.}
We discover the middle layer of the APC family, which is farthest from the surface feature, achieves the highest bound against both labeled data and future view.
More importantly, a consistent trend can be found across all three columns within each model, suggesting that the \um~ can be used to select the target layer for feature extraction.
A similar study on MLM models is carried out in Fig~\ref{fig:layer}.
Conversely, we find the trend of the lower bound of \smi~and \umi~to match at the early and last layers of the model.
This shows that different self-supervised learning methods result in different representation patterns as suggested by existing study~\cite{pasad2021layer,pasad2023comparative}.
Nevertheless, we note that using the last layer with \umi~lower bound is a robust option for comparing different models at the same size as it consistently matches the downstream performance trend as shown in Table~\ref{tab:mlm} and Table~\ref{tab:apc}.

\vspace{3pt}
\noindent\textbf{Robustness for checkpoint selection.}
In Figure~\ref{fig:mask} and Figure~\ref{fig:cluster}, we showcase the robustness of \umi~lower bound by using it to evaluate DinoSR at different pre-training steps given that checkpoint selection is sometimes a hard problem for self-supervised methods in practice.
To simulate the training scenario, this part is conducted on the validation set only by splitting it into half for fitting $f_\text{cluster}$ and MLP $\probe$, using the remaining part to estimate MI.

For MLM-based methods, the masking ratio is an important hyper-parameter that controls the view $X_a$ used for training.
Prior works have applied different masking ratios varying between 65\% to 80\%\footnote{In practice, these methods randomly sampled 6.5\% to 8\% of input frames to apply mask spanning 10 consecutive frames. This results in a lower masking ratio than expected since spans might overlap each other. For our evaluation, the expected masking ratio is precise since masks are not overlapping.}.
While lower masked ratios result in higher absolute MI estimation, we find \umi~lower bound robust to the choice of masking ratio, providing a consistent pattern as the training continues.

Finally, as shown in Figure~\ref{fig:cluster}, we observe similar trends of \umi~lower bound despite varying the number of clusters.
These findings point out a new path for evaluating self-supervised speech models during pre-training as computing \umi~lower bound requires little computations~\footnote{Our method required 2 forward passes on the validation set. As a reference, each estimation with MLP takes less than 5 minutes on CPU for fitting $f_\text{cluster}$ and $\probe$ (can be further sped up by leveraging GPU), which is considerably short compared to the runtime of self-supervised methods itself.}.

%% file: tables/apc.tex
\begin{table*}[t!]
\begin{minipage}{0.4\textwidth}
\small
    \caption{Results of APC family on LibriSpeech test-clean. MLP is used as $q_\theta$ for bounding conditional entropy. LB stands for low-bound. The layer-wise PER are evaluated following the protocol of \cite{chung2020generative,yang2022autoregressive,yeh2022autoregressive}. 
    All models are trained and evaluated with the same time-shift of 60ms.
    }
    \label{tab:apc}
    \centering
    \resizebox{1.0\linewidth}{!}{
    \begin{tabular}{lccc}
    \toprule
    & \smi~LB  & \umi~LB & PER  \\
    &  Layer 1/2/3 & Layer 1/2/3 & Layer 1/2/3 \\
    \midrule
    APC & 3.0~/~3.6~/~3.3 & 2.3~/~2.8~/~2.3 & 25.3~/~23.8~/~31.4 \\
    VQ-APC & 3.4~/~3.6~/~3.1 & 2.8~/~3.0~/~2.9 & 25.4~/~22.7~/~28.4 \\
    Co-training & 3.3~/~3.6~/~3.2 & 2.7~/~3.0~/~2.5  & 27.1~/~21.0~/~27.1\\
    \bottomrule
    \end{tabular}
    }
\end{minipage}
\begin{minipage}{0.29\textwidth}
\captionsetup{width=\linewidth}
    \centerline{\includegraphics[width=\linewidth]{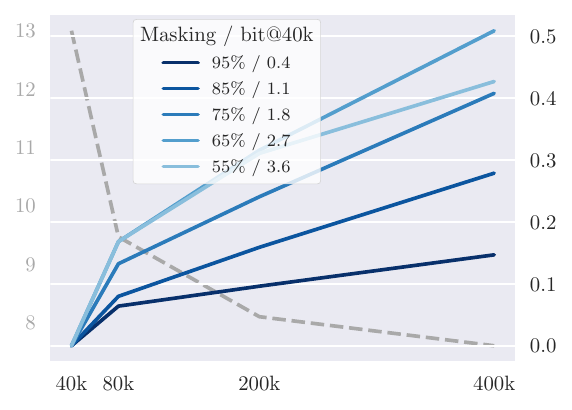}}
    \vspace{-8pt}
     \caption{\small Varying masking ration for bounding \umi~on DinoSR. X-axis: pre-training step. Y-axis: the improvement of lower-bound compared to 40k steps (bit; right) and the fine-tuning performance (gray dashed line; WER; left).}
    \label{fig:mask}  
\end{minipage}
\begin{minipage}{0.02\textwidth}
\end{minipage}
\begin{minipage}{0.29\textwidth}
    \captionsetup{width=\linewidth}
    \centerline{\includegraphics[width=\linewidth]{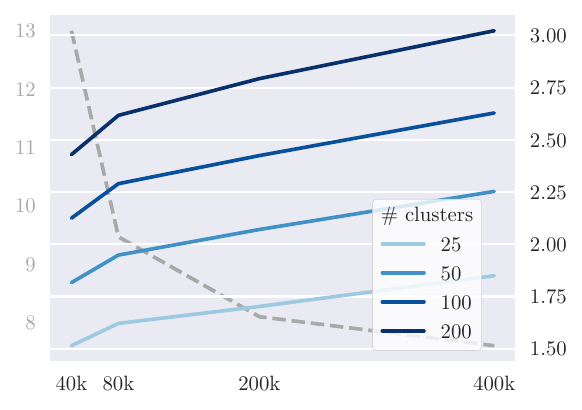}}
    \vspace{-8pt}
    \caption{\small Varying $f_\text{cluster}$ cluster size for bounding \umi~on DinoSR. X-axis: pre-training steps. Y-axis: \umi~lower-bound (bit; right) and the fine-tuning performance (gray dashed line; WER; left).}
    \label{fig:cluster}
\end{minipage}
\vspace{-10pt}
\end{table*}

%% file: sections/conclusion.tex
\section{Conclusion}

In this paper, we revisited self-supervised learning of speech representation from a mutual information point of view.
We provided two different MI metrics and showed their lower bounds can be used to evaluate self-supervised models without heavy computations, especially for the \umi~lower bound that is designed in a self-supervised manner.
We checked the robustness of these metrics to demonstrate the potential of applying them to different pre-trained models in practice.
However, we also note that this work focused on examining the content of speech as we only considered recognition tasks in our experiment.
An interesting future direction is to explore the non-content information encoded in speech representations.